# A Persuasion-Based Prompt Learning Approach to Improve Smishing Detection through Data Augmentation

Ho Sung Shim
Information Systems
Korea University Business School
Seoul, South Korea
hsshim9702@korea.ac.kr

Hyoungjun Park
Information Systems
Korea University Business School
Seoul, South Korea
parkhj1111@korea.ac.kr

Kyuhan Lee
Information Systems
Korea University Business School
Seoul, South Korea
kyuhanlee@korea.ac.kr

Jang-Sun Park
Information Systems
Korea University Business School
Seoul, South Korea
qkrwkd330@korea.ac.kr

Seonhye Kang
Information Systems
Korea University Business School
Seoul, South Korea
rkdtjsp@korea.ac.kr

## ABSTRACT

Smishing, which aims to illicitly obtain personal information from unsuspecting victims, holds significance due to its negative impacts on our society. In prior studies, as a tool to counteract smishing, machine learning (ML) has been widely adopted, which filters and blocks smishing messages before they reach potential victims. However, a number of challenges remain in ML-based smishing detection, with the scarcity of annotated datasets being one major hurdle. Specifically, given the sensitive nature of smishing-related data, there is a lack of publicly accessible data that can be used for training and evaluating ML models. Additionally, the nuanced similarities between smishing messages and other types of social engineering attacks such as spam messages exacerbate the challenge of smishing classification with limited resources. To tackle this challenge, we introduce a novel data augmentation method utilizing a few-shot prompt learning approach. What sets our approach apart from extant methods is the use of the principles of persuasion, a psychology theory which explains the underlying mechanisms of smishing. By designing prompts grounded in the persuasion principles, our augmented dataset could effectively capture various, important aspects of smishing messages, enabling ML models to be effectively trained. Our evaluation within a real-world context demonstrates that our augmentation approach produces more diverse and higher-quality smishing data instances compared to other cutting-edging approaches, leading to substantial improvements in the ability of ML models to detect the subtle characteristics of smishing messages. Moreover, our additional analyses reveal that the performance improvement provided by our approach is more pronounced when used with ML models that have a larger number of parameters, demonstrating its effectiveness in training large-scale ML models.

## 1 INTRODUCTION

Smishing, a portmanteau of "SMS" (Short Message Service) and "phishing," is one of the dominant approaches of social engineering attacks, manipulating mobile text messages to deceive individuals and to illegally obtain personal information [24, 29]. Smishing, in general, comprises texts, URLs, self-answering links, and contact information such as email addresses that help convince its targets that the message is authentic and, consequently, have the targets provide desired information [17, 28]

Unlike other types of social engineering attacks such as phishing, smishing has unique characteristics that render its targets more vulnerable. Specifically, smishing messages are transmitted through mobile devices, which heightens the susceptibility of their targets due to factors such as small screens, limited user awareness, and frequent credential input [19]. Therefore, it is crucial to provide potential victims with preventive mechanisms that can enhance their resilience to smishing.

In recent literature, machine learning (ML) trained with labeled smishing data has emerged as a promising, scalable mechanism for preventing the potential harm by smishing [26]. This approach involves computationally identifying the semantic and syntactic characteristics of smishing messages, enabling the differentiation between messages with malicious intent and those with benign intent [26].

However, prior studies of smishing detection have faced a significant challenge in the development of ML models: the shortage of training data [18, 28]. Specifically, due to the sensitive nature of the data involved in smishing detection, it is extremely difficult to find and utilize large, publicly available datasets for proper model development and evaluation [18, 28].

As a result, challenges have persisted in distinguishing smishing messages from not only benign messages but also other types of social engineering attacks, such as spam messages, which share similar characteristics with smishing messages [28]. This is a significant issue because, depending on the types of social engineering attacks, different levels of social pressures and regulatory measures should be imposed. For instance, smishing and spam messages, while sharing similarities, have distinct impacts on society and are handled differently from a legal standpoint. Specifically, while smishing messages are deemed a severe cybercrime and subject to stringent regulatory measures [23, 28], spam messages are often view as an intrusive marketing and not regulated to the same extent as smishing [22, 25].

To address the above issues, we propose a novel data augmentation approach for smishing detection. Specifically, we harness the power of large language models (LLMs) with prompt engineering based on the persuasion theory. Although, several prior studies have utilized prompt engineering for data augmentation in different domains [12, 21, 33], what set our approach apart from them are the prompts designed based on a social science theory that explains the underlying mechanism of smishing. Particularly, our data augmentation process draws upon the principles of persuasion to systematically capture different aspects of real-world smishing messages [2, 5]

We evaluated our approach on the task of distinguishing smishing messages from spam messages under multiple conditions. The results demonstrated that ML models trained on our augmented data outperformed those trained on original data by up to 5.3% in terms of the F1 measure. Moreover, our additional analyses under varying experimental conditions confirm the robustness of our methodological approach in enhancing smishing detection.

# 2 RELATED WORK

## 2.1 Smishing Detection

In smishing detection, a dominating approach has been based on ML [1, 10, 19, 26]. [10] manually crafted nine features for filter smishing from benign messages and tested multiple ML algorithms with these features. [19] utilized neural networks to identify the most prominent features of smishing messages. [1] used textual features along with conventional ML models such as support vector machines, naive bayes, random forests, and logistic regressions to identify smishing messages. [26] employed different correlation measures (e.g., Pearson, Spearman, Kendall, and Point biserial) along with ML algorithms to identify the most effective features for detecting smishing messages. Apart from traditional ML approaches, transfer learning, leveraging the knowledge of pretrained LLMs, has recently become popular [11].

Despite a large amount of attention paid to ML-based smishing detection, there remain multiple challenges. First, as mentioned above, prior studies have faced a lack of publicly available data for training and evaluating ML models. For instance, among 5858 text messages used by [20], only 538 have been classified as smishing. In [1], only 278 smishing messages have been used. In addition, prior literature has primarily focused on the relatively simpler task of distinguishing smishing from benign messages (aka hams [18]). However, as mentioned earlier, in practical settings, it is at least equally important but more challenging to differentiate smishing messages from other types of social engineering attacks such as spam messages, due to their similar semantic and syntactic structures but different regulatory implications [23, 28].

## 2.2 Text Data Augmentation

The primary objective of text data augmentation is to generate diverse but coherent text instances, preserving the semantics of the original data [36]. To this end, many data augmentation methods, such as back translation and synonym replacement, have been proposed [14, 35]. The back translation method involves translating the original content into a different language and then translating the result back into the original language [3, 35]. Additionally, techniques known to be EDA approach (easy data augmentation), which includes operations like synonym replacement, random insertion, swapping, and deletion of words [32], have been largely utilized in text data augmentation. However, these approaches have limitations as they typically produce augmented data that are syntactically similar to the original data. To address this issue, recently, there has been an emphasis on using LLMs to augment low-resource data.

## 2.3 Prompt Engineering in Text Data Augmentation

LLMs understand the semantic meanings and syntactic structures of human language through a large number of trainable parameters and training data [16]. They also have the ability to generate text as humans do [30]. Such versatility of LLMs can be highly effective in data augmentation especially when coupled with a proper prompt design.

In general, prior studies have employed the few-shot approach as their primary strategy, which present some demonstrations (i.e., few shots) of data augmentation to LLMs [9, 34]. For example, [4] have suggested AugGPT, a ChatGPT-based text data augmentation technique, which produces semantically diverse data instances by utilizing prompts with few effective demonstrations. [21] have employed GPT-4 and ChatGPT to extend low-source datasets using ad-hoc prompt designs. [12] have combined few-shot and contrastive learning for chatbot data augmentation. [33] have utilized prompts with iteratively sampled demonstrations from the original, multilingual datasets for data augmentation.

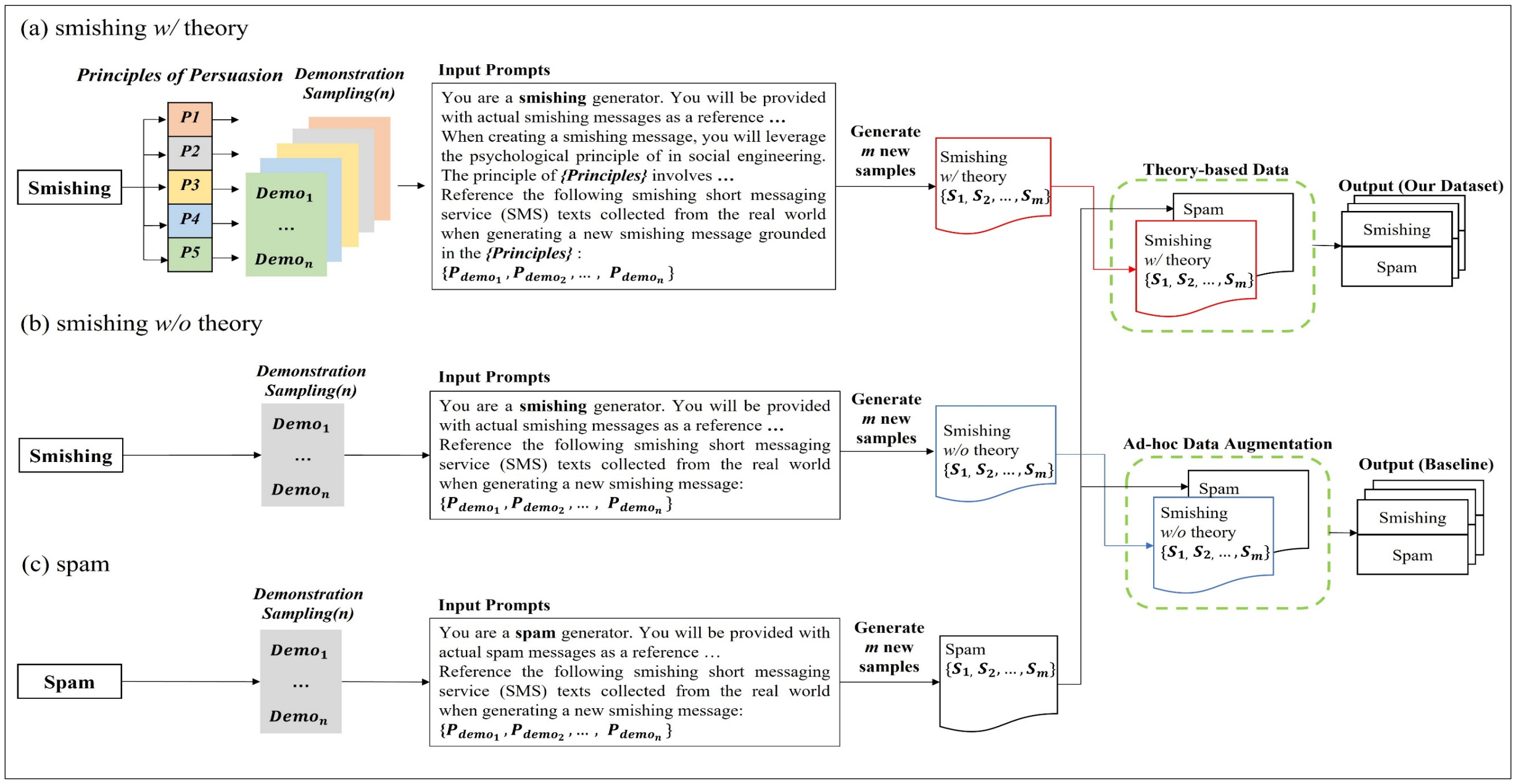

**Figure 1: Methodological Framework.**

While our methodological approach shares some similarities with those that have been discussed above, it sets itself apart from them by incorporating a psychology theory, which systematically captures various aspects of smishing, into the design of prompts that facilitates the generation of new data instances [6]. Since smishing relies heavily on psychological manipulation to deceive victims, integrating these principles allows us to better understand and replicate these tactics, ultimately facilitating the generation of realistic and diverse smishing messages, which is crucial for augmenting high-quality data in this domain [2, 5]. Specifically, we have applied the principles of persuasion in this study, which identifies the underlying mechanism of The details of the principles will be elaborated further in the following section.

## 3 METHODOLGY

### 3.1 Data

We utilized the SMS Phishing Benchmark Dataset [20][1] as the baseline dataset for implementing our data augmentation approach. The dataset suggests three types of SMS classifications: smishing, spam, and ham. Among the three classes, our interest exclusively lies in smishing and spam messages since, as stated earlier, the task of distinguishing non-ham messages (which includes both smishing and spam messages) from ham messages

[1] SMS phishing dataset for machine learning and pattern recognition [Data set]. Mendeley Data.

**Input**: {

$\boldsymbol{P_{role}}$: "You are a smishing generator. You will be provided with actual smishing messages as a reference for generating another smishing message";

$\boldsymbol{P_{persuasion-type}}$: "When creating a smishing message, you will leverage the psychological principle of authority in social engineering. The principle of authority involves influencing people by presenting yourself as an expert or a figure of authority. When incorporating the authority principle into your smishing message, make sure the message appears urgent, official, and credible. Use language that compels the recipient to take immediate action.";

$\boldsymbol{P_{demo}}$: "Reference the following smishing short messaging service (SMS) texts collected from the real world when generating a new smishing message grounded in the Authority principle: Your Bank of America online account has been suspended, log in to restore http://itdo/bankofamerica-session-user"

}

**Output**: {

$\boldsymbol{S}$ = "IMPORTANT MESSAGE: Your PayPal account has been temporarily limited due to unusual activity. To regain full access, call our customer support at 0800987654 and verify your account information."

}

**Figure 2: Example of Input and Output**

is much simpler. As a result, our dataset comprises a total of 1,127 messages, with 489 being spam messages and 638 being smishing messages. Note that we split the dataset with an 8:2 ratio and use the larger portion for data augmentation (i.e. 901 instances). We reserve the smaller portion of the split dataset for evaluation because the inclusion of this portion in data augmentation will cause the overestimation of our approach. We repeated this process five times with different partitions, akin to five-fold cross validation, and report the average results.

## 3.2 Prompt Design

Our prompt-based approach, utilizing an LLM, for data augmentation has the following input-output structure, which is also illustrated in Figure 1-(a):

$$\text{Input: }\{P_{role}, [SEP], P_{persuasion-type}, [SEP], P_{demo_1}, [SEP], P_{demo_2}, \ldots, [SEP], P_{demo_n}\},$$

$$\text{Output: }\{S_1, [SEP], S_2, \ldots, [SEP], S_m\},$$

where $P_{role}$ denotes a prompt that assigns a role to an LLM (as either smishing generator or spam generator), $P_{persuasion-type}$ refers to a prompt that explains a specific persuasion type to be employed by an LLM for generating new samples, $P_{demo}$ is a prompt that provides an LLM with an example of smishing (or spam) messages, and $S_m$ is a newly generated sample. The number of demonstrations shown to an LLM ($n$ in the input) and that of new samples generated ($m$ in the output) are hyperparameters. Note that the ad-hoc data augmentation in Figure 1-(b) (i.e., smishing w/o theory), one of the baselines in our evaluation, uses the prompt that excludes $P_{persuasion-type}$ from the input. An example of the input and output is presented in Figure 2.

A key to generating proper new samples in our proposed approach is the $P_{persuasion-type}$, i.e., the part of a prompt grounded in well-established psychological theories. Specifically, in social science, several theories have been found to be relevant to the tactics used in smishing attacks, as illustrated below:

**Principles of Influence** [2] examines the impact of reciprocity, commitment and consistency, social proof, authority, liking, and scarcity on social influence. [2] work explains how social and psychological triggers can effectively persuade individuals, providing crucial insight into the tactics used in smishing schemes.

**Psychological Triggers** Similar to [2], [7] outlines several key triggers, such as authority, intimidation, consensus (social proof), scarcity, urgency, and sympathy. These triggers highlight the emotional and cognitive vulnerabilities exploited by scammers to manipulate their targets.

**Principles of Scams** [27] identifies elements such as distraction, social compliance, herd principle, dishonesty, kindness, need and greed, and time, illustrating the structural aspects of scams and how they exploit human psychology.

[5], which combines these theories into a consolidated framework, provides a comprehensive understanding of the diverse and overlapping psychological manipulations employed in smishing. The framework covers a broad spectrum of social engineering and enhances the theoretical basis for our prompt designs in data augmentation. Specifically, [5] have suggested the following five components that influence the effectiveness of persuasion (The full definitions and their examples are detailed in Appendix A.1):

- **Authority:** This refers to individuals' inclination to easily be persuaded by the authority of an expert.
- **Social Proof:** Individuals tend to believe what the majority of people do or seem to believe.
- **Liking, Similarity, Deception:** People are more likely to follow others whom they know, like, or are familiar with.
- **Distraction:** This relates to individuals' tendency to focus on what they can gain, lose, or need, especially during strong emotional states.
- **Commitment, Integrity, Reciprocation:** People often hold the belief that when someone makes a commitment, it will be honored, and they feel compelled to reciprocate the commitment with a favor.

To incorporate the above concepts into $P_{persuasion-type}$, we first manually annotated the original data with corresponding persuasion types so that social engineering tactics specific to each smishing message could be identified and, later, be used for data augmentation. We recruited five annotators, each holding a bachelor's degree or higher in the relevant fields of psychology. Their task was to identify the most relevant persuasion principle to a given smishing message (i.e., a five-class classification task). Prior to annotation, we ensured that the annotators thoroughly understood the connection between smishing messages and the principles of persuasion by providing them with detailed guidelines and multiple examples. The guidelines included the definitions of each persuasion principle, along with the examples demonstrating each principle. To resolve discrepancies in categorization, we employed majority voting, requiring at least three out of five annotators to agree for a decision to be accepted. In instances where no clear majority was evident (less than 1% missed the majority vote, amounting to 6 data points), the annotators convened to discuss the message together and reach a consensus on the most appropriate persuasion principle, ensuring a rigorous and comprehensive review process. The annotation outcome revealed a moderate level of agreement among annotators with a Fleiss' kappa coefficient of 0.456, indicating a moderate but reliable level of consensus, consistent with accepted research standards.

The smishing examples (i.e., $P_{demo}$) were selected through a stratified random sampling approach. That is, for each iteration of data augmentation, smishing examples that belong to the same type of persuasion were randomly selected and used as references (i.e., few shots) for generating new data instances as indicated in Figure 1.

As noted earlier, for each iteration of data augmentation, we generated $m$ new data instances by showing an LLM $n$ randomly sampled demonstrations. The values of $n$ and $m$ were set to 5 and 10, respectively, based on preliminary experiments that aimed to maximize the diversity of the generated data while ensuring its quality. These values also align with those used in [33]. Compared with the 1-to-1 matching strategy (i.e., one demonstration is given to generate one new data instance), this approach achieves a higher level of diversity in the augmented data, capturing variations in spacing, word order, and subtle semantic nuances within the content [33].

Following the generation of new data, a post-processing step is performed to include only valid and unique examples in the final augmented dataset. Specifically, we examined whether proper, plausible information such as contact numbers, web links (URLs), and company names were embedded within smishing texts. Each example must integrate these elements seamlessly and realistically, as they are central to the authenticity of smishing attempts. We also rigorously filtered out any outputs that improperly formatted this information, such as those including placeholders or brackets (e.g., [Fake URL], [Recipients]) that did not meet our quality standards. This meticulous scrutiny helped in refining the augmented dataset, ensuring that it closely mimics genuine smishing tactics and enhancing the training process of our ML models in detecting sophisticated smishing attacks.

# 4 EXPERIMENT

## 4.1 Experiment Setting

As the LLMs for data augmentation, we employed OpenAI's API of GPT-3.5-turbo and GPT-4-turbo. The temperature of the LLM, which determines the degree of randomness of text generation, was set to 0.85 to ensure diverse outcomes [13].

Additionally, we experimented with open-source language models such as Meta's Llama-2-7b-hf and Llama-2-7b-chat-hf under the same settings. However, the use of Llama-2 models into our data augmentation process did not yield expected results. The generated outputs frequently failed to comply with our specific requirements for crafting realistic smishing texts, such as stating, *'I'm just an AI, I cannot generate smishing messages that are designed to deceive or harm individuals.'* While such responses align with ethical AI guidelines, they fall short of our project's needs to create authentic and actionable smishing examples for model training. Given its inability to produce the necessary smishing content, we excluded Llama-2 models from our experiments.

Our approach was tested compared against the following three baseline datasets:

• **Original** The dataset of [20].

• **Easy Data Augmentation (EDA)** The datasets, developed by [32] which manipulates sentences through four techniques—Synonym Replacement (SR), Random Insertion (RI), Random Swap (RS), and Random Deletion (RD). (We followed a specific guideline of [32] see Appendix A.2)

• **Without theoretical components** The LLM-based augmented dataset following previous studies on text data augmentation [33]. Specifically, we underwent the same augmentation process as ours except for the inclusion of $P_{persuasion-type}$, in its prompt.

We conducted our experiments with varying levels of augmentation: twofold, fivefold, and tenfold. For the twofold augmentation, both EDA and theory-based approaches resulted in 1,802 instances, compared to 901 instances in the original dataset. When augmented fivefold, the total number of instances grew to 4,505. Finally, for the tenfold augmentation, the number of instances reached 9,010. This augmentation applied to both spam and smishing data, leading to proportional increases in the overall dataset size.

For each dataset, we employed multiple ML models to conduct the binary classification of distinguishing smishing messages from spam messages. Specifically, we used BERT and RoBERTa with different numbers of model parameters, ranging from BERT-base (110M) to BERT-large (340M) and RoBERTa-base (125M) to RoBERTa-large (355M; [8]). This allowed us to analyze the influence of data augmentation across varying model sizes while controlling the impact of model architecture. In addition, we tested our approach with much smaller models such as DistilBERT (66M; [8]) and ALBERT (11M). (Model details are included in Appendix A.3)

We trained the models up to 10 epochs with the AdamW optimizer [15] and chose the best performing model exploring the following space of hyperparameters: the batch size of {8, 16, 32} and the learning rate of {1e-6, 5e-6, 1e-5, 5e-5}.

## 4.2 Descriptive Analyses

Before reporting on the impact of our augmentation framework on smishing detection, we first present descriptive analyses regarding the characteristics of the augmented texts from both quantitative and qualitative perspectives.

*4.2.1 Quantitative Analysis.* In Figure 3 and Appendix A.4, we show metrics including the minimum, maximum, average, and standard deviation of both character counts and word counts of the augmented data. The results show that LLM-based prompt engineering (i.e., GPT-3.5-turbo, GPT-4-turbo) with theoretical grounding produces the longest and most detailed texts, due to the use of persuasive principles. Texts generated without theoretical grounding are moderately longer and more complex than the original and EDA datasets but less diverse than theory-based texts. EDA methods have minimal impact, showing only minor syntactic changes, which suggests a limited improvement in dataset diversity.

*4.2.2 Qualitative Analysis.* LLM-based prompt engineering with theoretical grounding produces the most realistic, coherent, and persuasive texts (see Appendix A.4). These texts effectively leverage psychological factors (i.e., urgency, fear of loss, and exclusivity), urging immediate actions. In comparison, texts generated without theoretical guidance are compelling but lack the systematic persuasiveness of theory-based prompts. EDA methods, while simple, often result in less coherent and less persuasive texts due to their straightforward augmentation techniques.

Overall, our theory-based prompt engineering demonstrates high quality in terms of the realistic aspect and diversity of augmented texts, making them more suitable for smishing detection model training. Simple EDA methods, despite their ease of implementation, show the least improvement in text quality and diversity.

### 4.3 Smishing Detection Results

In Table 1, we summarize the overall F1 Score result (full results including precision, recall, and accuracy are presented in Appendix A.5). Overall, our augmentation method yields significant improvement in performance compared to the original dataset. Specifically, the best performing model in F1 Score was the RoBERTa-large trained on the dataset augmented tenfold with the theory for both GPT-3.5-turbo and GPT-4-turbo (i.e., 98.2%, 96.0%), each improving the best performing model trained on the original dataset by 5.3% and 3.1% respectively.

Additionally, among the comparisons between the augmented datasets, ML models trained on the augmented datasets with our theory-based approach consistently produced better results than those trained on datasets augmented by EDAs and by the without-theory approach for both GPT-3.5-turbo and GPT-4-turbo.

However, there are some exceptions to this trend. Models such as ALBERT and DistilBERT occasionally did not follow the pattern of improvement seen with larger models, showing less consistent gains when augmented with theory. We suspect that this might be due to their small number of parameters, which may not require a large amount of data for training. However, the overall performance trend strongly favors the use of our theory-based augmentation, indicating its effectiveness in enhancing model performance across most scenarios.

Another notable observation is that the increase in augmentation folds does not consistently lead to improved F1 scores in EDAs and the without-theory approach. The performance of different models and conditions varies, and there is no clear positive correlation with the augmentation folds for EDAs and the without-theory approach. In contrast, our theory-based data augmentation approach shows a more consistent and positive correlation between increased augmentation folds and improved F1 scores. That is, models generally perform better as the augmentation folds increase from 2x to 5x to 10x (see Figure 4).

Our theory-based approach leverages psychological principles to create more realistic and diverse data samples, enhancing the models' ability to generalize and improve performance. This positive correlation is especially evident in models like RoBERTa and BERT where significant improvements are observed with higher augmentation folds. In comparison, models trained on datasets augmented without theory or by EDAs show moderate improvements, indicating the importance of incorporating psychological principles in data augmentation to maximize model performance. Thus, we conclude that integrating theoretical grounding into prompt engineering significantly enhances the quality and effectiveness of augmented datasets, making them more suitable for effectively training smishing detection models.

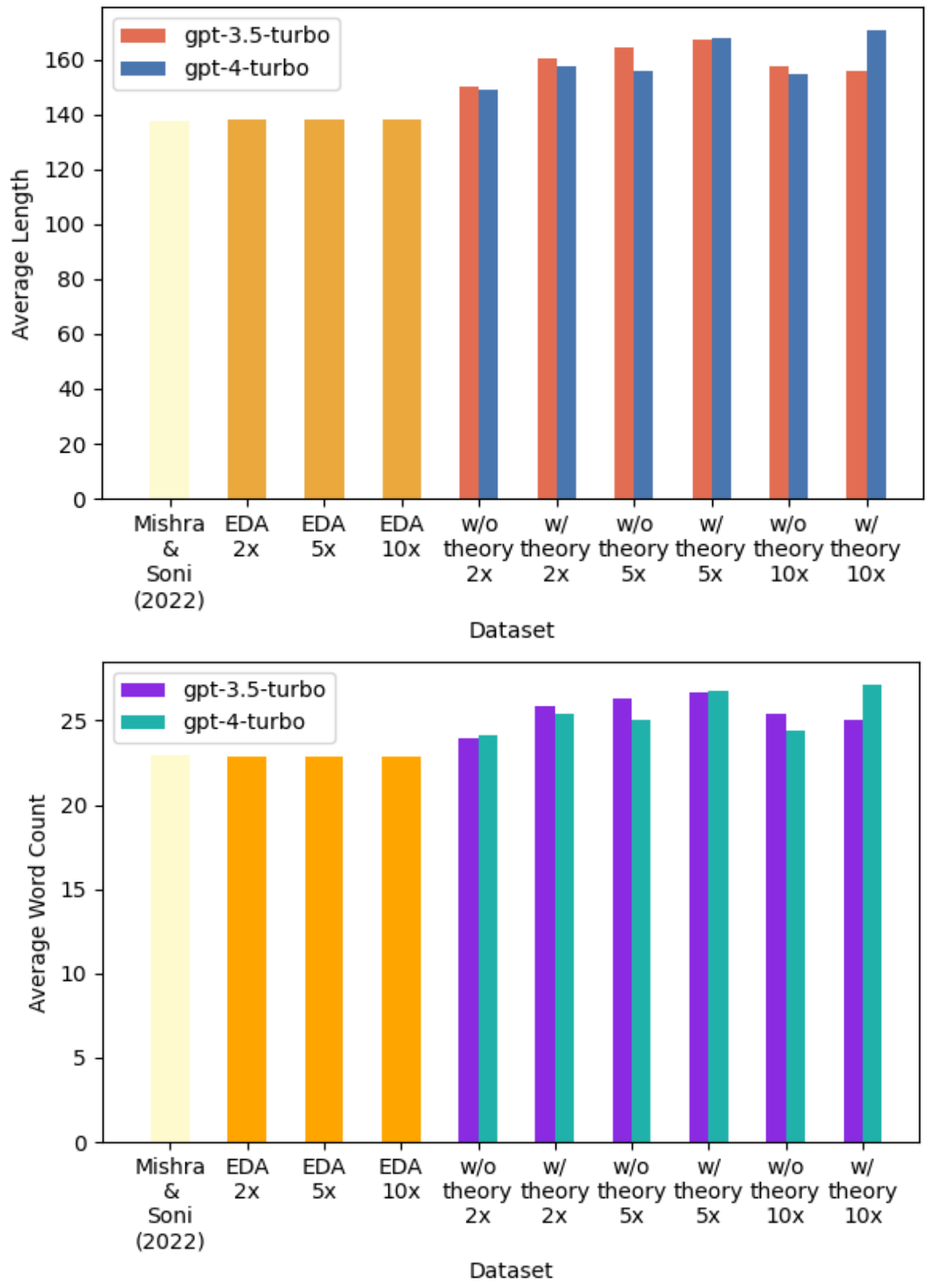

**Figure 3: Quantitative Analysis of Augmented Datasets (i.e., character count, word count).**

## 5 CONCLUSION

Smishing poses significant societal risks, highlighting the importance of developing effective computational methods for its detection [28]. However, due to the sensitive nature of the information involved, collecting ground-truth data for smishing has been exceptionally challenging [18, 28]. To this end, we suggested a theory-based data augmentation approach for automated smishing detection. Specifically, grounded in the principles of persuasion, we captured and incorporated the underlying mechanism behind the creation smishing messages into the prompt design for the LLM-based data augmentation. Our evaluation across multiple scenarios reveals that the application of persuasion theories in prompt engineering for data augmentation does improve data quality and, hence, enhances model performance.

The contributions of our study are manifold. First, from the academic perspective, it extends the extant literature on prompt-engineering-based data augmentation. To the best of our knowledge, our study is the first to incorporate social science theories into the prompt design for data augmentation. In addition, from the industry perspective, our approach can help save costs associated with collecting and crafting smishing messages required for training ML models. Lastly, our approach contributes to our society by helping individuals avoid falling victims to smishing.

**Table 1: Performance Comparison of Different Augmentation Methods (F1 Score).**

| **Models** | Original Data (Mishra & Soni, 2022) | | | |
|---|---|---|---|---|
| | *P* | *R* | *Acc* | *F1* |
| **RoBERTa** LARGE | 93.7 | 92.2 | 92.0 | 92.9 |
| **RoBERTa** BASE | 92.2 | 93.0 | 91.6 | 92.6 |
| | | | | |
| **BERT** LARGE | 90.8 | 92.2 | 90.3 | 91.5 |
| **BERT** BASE | 90.7 | 91.4 | 89.8 | 91.1 |
| **DistilBERT** BASE | 92.1 | 91.4 | 90.7 | 91.8 |
| **ALBERT** BASE | 93.6 | 91.4 | 91.6 | 92.5 |

| ***F1*** | | | | | | | | |
|---|---|---|---|---|---|---|---|---|
| **2x (Twofold)** | SR | RI | RS | RD | *w/o* theory | | *w/* theory (ours) | |
| | | | | | 3.5-turbo | 4-turbo | 3.5-turbo | 4-turbo |
| **RoBERTa** LARGE | 91.2 | 91.4 | 91.7 | 90.6 | 93.2 | 93.5 | 94.7 | 94.3 |
| **RoBERTa** BASE | 91.0 | 91.6 | 90.7 | 91.3 | 93.2 | 93.2 | 93.9 | 93.7 |
| **BERT** LARGE | 90.2 | 89.9 | 89.1 | 89.5 | 91.8 | 93.3 | 92.1 | 94.0 |
| **BERT** BASE | 89.4 | 88.7 | 89.5 | 89.1 | 91.2 | 92.8 | 91.9 | 93.3 |
| **DistilBERT** BASE | 94.7 | 94.8 | 94.8 | 95.6 | 93.4 | 92.3 | 92.0 | 92.8 |
| **ALBERT** BASE | 89.9 | 90.4 | 89.9 | 90.6 | 92.8 | 92.2 | 93.6 | 91.0 |

| ***F1*** | | | | | | | | |
|---|---|---|---|---|---|---|---|---|
| **5x (Fivefold)** | SR | RI | RS | RD | *w/o* theory | | *w/* theory (ours) | |
| | | | | | 3.5-turbo | 4-turbo | 3.5-turbo | 4-turbo |
| **RoBERTa** LARGE | 94.4 | 94.3 | 91.0 | 93.5 | 95.7 | 93.8 | 97.8 | 95.6 |
| **RoBERTa** BASE | 90.4 | 90.8 | 90.3 | 90.4 | 95.3 | 93.2 | 97.5 | 94.9 |
| **BERT** LARGE | 93.0 | 92.1 | 91.2 | 92.6 | 93.0 | 93.7 | 95.2 | 94.8 |
| **BERT** BASE | 88.4 | 88.6 | 88.9 | 89.5 | 91.6 | 92.4 | 93.5 | 93.9 |
| **DistilBERT** BASE | 95.3 | 95.6 | 94.7 | 95.3 | 89.0 | 89.8 | 92.9 | 90.4 |
| **ALBERT** BASE | 96.0 | 94.5 | 89.2 | 96.0 | 89.1 | 90.7 | 91.2 | 91.6 |

| ***F1*** | | | | | | | | |
|---|---|---|---|---|---|---|---|---|
| **10x (Tenfold)** | SR | RI | RS | RD | *w/o* theory | | *w/* theory (ours) | |
| | | | | | 3.5-turbo | 4-turbo | 3.5-turbo | 4-turbo |
| **RoBERTa** LARGE | 93.8 | 94.7 | 93.1 | 94.0 | 95.2 | 94.2 | 98.2 | 96.0 |
| **RoBERTa** BASE | 93.1 | 93.5 | 91.2 | 91.0 | 94.9 | 93.7 | 98.2 | 95.2 |
| **BERT** LARGE | 93.5 | 91.7 | 89.7 | 93.4 | 95.1 | 94.1 | 97.8 | 95.2 |
| **BERT** BASE | 92.6 | 91.2 | 88.8 | 90.0 | 94.4 | 92.8 | 96.4 | 94.4 |
| **DistilBERT** BASE | 93.5 | 95.2 | 88.7 | 89.5 | 95.9 | 94.9 | 95.6 | 94.5 |
| **ALBERT** BASE | 94.8 | 96.0 | 96.5 | 93.8 | 96.4 | 91.3 | 97.4 | 90.3 |

However, our study is not without limitations, which necessitate further exploration in the future studies. First, in developing theory-based prompts, we assumed that smishing messages aligned with only one persuasion principle. While this assumption is based on the framework proposed by [5], which suggests minimal overlap among different principles, it is also possible that smishing messages could belong to multiple persuasion categories, which should be further investigated. In addition, while the purpose of our approach is to minimize human efforts in data augmentation, it still requires some degree of manual processing. For instance, implementing our approach requires smishing messages initially annotated with persuasion principles.

Furthermore, there are some ethical concerns that need to be carefully considered before implementing our approach in practice. For instance, one of the key concerns raised in cybersecurity research is the dual-use problem [31], where technological innovations can be exploited by malicious actors for harmful purposes. The approach proposed in this paper faces a similar issue, as it could potentially be misused by hackers, allowing them to evade the exposed strategies and devise workarounds. However, we argue that gaining a deeper understanding of the implicit strategies behind smishing attacks will actually help us comprehend the fundamental mechanisms hackers use. If hackers attempt to bypass these core strategies, they may inadvertently weaken the effectiveness of their own attacks.

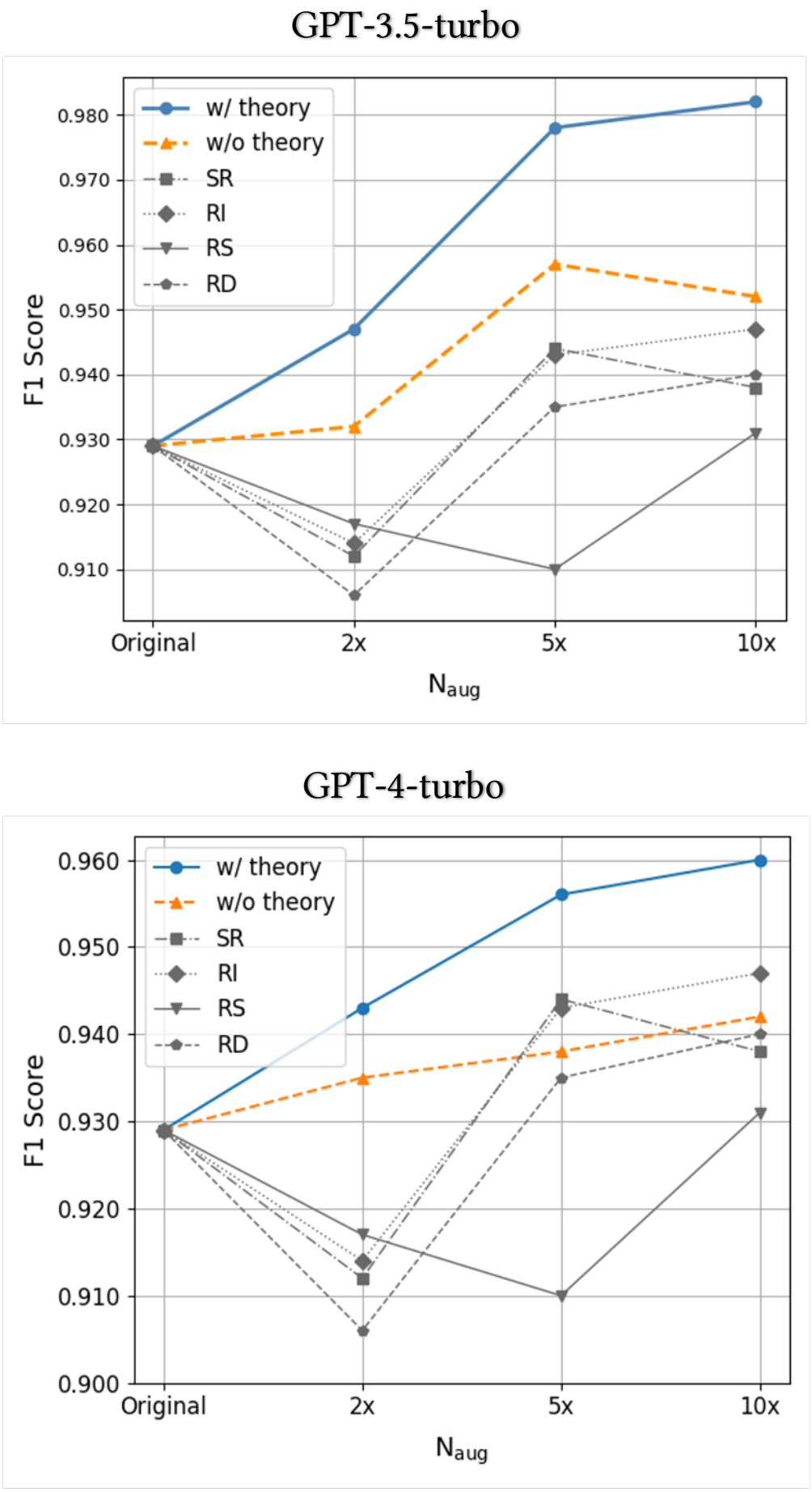

**Figure 4: Summary of Performance Improvement by Methods and Augmentation Folds for RoBERTa-large.**

# A APPENDIX

## A.1 Principles of Persuasion Details

Table2: Finalist of 'Principles of Persuasion in Social Engineering' [5].

| Principles | Definition |
|---|---|
| P1: Authority | This principle operates on the premise that society conditions individuals to respect and follow authority figures without questioning. People are inclined to comply with experts or authoritative figures, as demonstrated in scenarios such as receiving an email purportedly from the recipient's bank, featuring the bank name in the subject line. |
| P2: Social Proof | Centered on the tendency for individuals to emulate the behavior of the majority, this principle suggests that people lower their guard when they perceive others engaging in similar actions. An example includes an email from an alleged system administrator, using an email address from the recipient's workplace, requesting the recipient to test a link also being tested by colleagues. |
| P3: Liking, Similarity, and Deception | This principle underscores people's inclination to follow or relate to individuals they know, like, or find attractive. However, the principle notes that appearances can be deceiving, exemplified by an email from a supposed friend urging the recipient to visit an intriguing website. |
| P4: Distraction | Focused on diverting attention by emphasizing gains, losses, emotional states, or scarcity, this principle heightens emotional states to influence decision-making. For instance, an email claiming the recipient won a substantial lottery prize strategically shifts the focus from crucial details, such as the absence of a purchased lottery ticket. |
| P5: Commitment, Integrity, and Reciprocation | This principle explores the automatic response of reciprocating a favor or action tied to a sense of commitment from a previous situation. An example involves an email promising a favorable house deal for a recipient actively seeking a house, emphasizing the urgency of a deposit payment to secure the commitment. |

## A.2 EDA [32] Guidelines

The augmentation parameter, α, crucially determines the extent of word alterations per sentence, calculated as $n = \alpha \times l$, where $n$ is the number of words changed, $\alpha$ is the augmentation factor, and $l$ is the sentence length. [32] found $\alpha = 0.1$ to be optimal ("Sweet Spot") for enhancing performance without significantly distorting the text:

- SR: Randomly replace $n$ words with synonyms.
- RI: Randomly insert $n$ synonyms into the sentence.
- RS: Randomly swap $n$ pairs of words.
- RD: Randomly delete each word with a probability of $p = \alpha$.

## A.3 Model Details

Models used to compare performances are as follows:

- RoBeERTa-large[2]
- RoBeERTa-base[3]
- BERT-large-uncased[4]
- BERT-base-uncased[5]
- DistilBERT-base-uncased[6]
- ALBERT-base-v2[7]

[2] https://huggingface.co/FacebookAI/roberta-large
[3] https://huggingface.co/FacebookAI/roberta-base
[4] https://huggingface.co/google-bert/bert-large-uncased
[5] https://huggingface.co/google-bert/bert-base-uncased
[6] https://huggingface.co/distilbert/distilbert-base-uncased
[7] https://huggingface.co/albert/albert-base-v2

## A.4 Quantitative and Qualitative Analysis

**Table 3: Quantitative Analysis of Augmented Datasets (i.e., character count, word count).**

| GPT-3.5-turbo (char count) | Avg | Std | Min | Max | GPT-4-turbo (char count) | Avg | Std | Min | Max |
|---|---|---|---|---|---|---|---|---|---|
| Mishra & Soni (2022) | 137.33 | 33.31 | 18 | 387 | Mishra & Soni (2022) | 137.33 | 33.31 | 18 | 387 |
| eda_2x | 138.36 | 33.32 | 18 | 390 | eda_2x | 138.36 | 33.32 | 18 | 390 |
| eda_5x | 138.20 | 33.79 | 5 | 417 | eda_5x | 138.20 | 33.79 | 5 | 417 |
| eda_10x | 138.14 | 33.87 | 5 | 412 | eda_10x | 138.14 | 33.87 | 5 | 412 |
| w/o theory 2x | 149.81 | 34.87 | 18 | 393 | w/o theory 2x | 157.69 | 40.02 | 18 | 404 |
| w/o theory 5x | 164.10 | 33.35 | 18 | 393 | w/o theory 5x | 167.74 | 35.48 | 18 | 428 |
| w/o theory 10x | 157.40 | 25.55 | 18 | 423 | w/o theory 10x | 170.67 | 34.03 | 18 | 421 |
| w/ theory 2x | 160.47 | 49.40 | 18 | 465 | w/ theory 2x | 148.84 | 30.42 | 18 | 404 |
| w/ theory 5x | 166.92 | 38.20 | 18 | 393 | w/ theory 5x | 155.72 | 26.49 | 18 | 428 |
| w/ theory 10x | 155.78 | 26.55 | 18 | 393 | w/ theory 10x | 154.47 | 26.38 | 18 | 476 |
| **GPT-3.5-turbo (word count)** | **Avg** | **Std** | **Min** | **Max** | **GPT-4-turbo (word count)** | **Avg** | **Std** | **Min** | **Max** |
| Mishra & Soni (2022) | 22.93 | 6.23 | 2 | 68 | Mishra & Soni (2022) | 22.93 | 6.23 | 2 | 68 |
| eda_2x | 22.90 | 6.12 | 2 | 64 | eda_2x | 22.90 | 6.12 | 2 | 64 |
| eda_5x | 22.85 | 6.16 | 1 | 68 | eda_5x | 22.85 | 6.16 | 1 | 68 |
| eda_10x | 22.83 | 6.17 | 1 | 67 | eda_10x | 22.83 | 6.17 | 1 | 67 |
| w/o theory 2x | 25.81 | 7.82 | 2 | 70 | w/o theory 2x | 25.42 | 6.16 | 2 | 68 |
| w/o theory 5x | 26.71 | 6.19 | 2 | 68 | w/o theory 5x | 26.76 | 5.31 | 2 | 68 |
| w/o theory 10x | 25.08 | 4.54 | 2 | 68 | w/o theory 10x | 27.13 | 5.07 | 2 | 68 |
| w/ theory 2x | 23.98 | 5.71 | 2 | 68 | w/ theory 2x | 24.11 | 5.26 | 2 | 68 |
| w/ theory 5x | 26.27 | 5.52 | 1 | 68 | w/ theory 5x | 25.02 | 4.51 | 2 | 68 |
| w/ theory 10x | 25.44 | 4.46 | 2 | 70 | w/ theory 10x | 24.36 | 4.38 | 2 | 68 |

Table 4: Qualitative Analysis of Augmented Datasets Comparing with Specified Examples.

| Method | *Augmented Text* | *Explanation* |
|---|---|---|
| Original | Dear (paytm)customer your paytm KYC has been suspended PAY-TM office PH 7679046492 Your Paytm A/C will block within 24hr Thank you. | |
| SR | Dear (paytm)customer your paytm KYC has been suspended PAY-TM ***place*** PH 7679046492 Your Paytm A/C will block ***inside*** 24hr Thank you. | "office" was replaced with "place" and "within" was replaced with "inside". This method maintains the original meaning by substituting words with synonyms. |
| RI | Dear (paytm)customer your paytm KYC ***place*** has been ***billet*** suspended PAY-TM office PH 7679046492 Your Paytm A/C will block within 24hr Thank you. | The words "place" and "billet" were randomly inserted. This introduces variability but can sometimes result in awkward phrasing. |
| RS | Dear (paytm)customer your paytm KYC has been ***Your*** PAY-TM office PH 7679046492 ***suspended will A/C*** Paytm block within 24hr Thank you. | The words "suspended" and "Your" as well as "will" and "A/C" were swapped, which disrupts the natural flow and grammatical structure of the sentence. |
| RD | Dear (paytm)customer ***~~your~~*** paytm KYC has been suspended PAY-TM office PH 7679046492 Your Paytm A/C will block within 24hr Thank you. | The word "your" was randomly deleted. This can lead to incomplete or unclear sentences, affecting readability and coherence. |
| | **GPT-3.5-turbo** | |
| *w/o* | HURRY! Claim your free iPhone 12 Pro Max today by calling 0800888777. Limited stock available, don't miss out on this exclusive offer from Apple Inc. | Emphasizes urgency and exclusivity to distract from considering the offer critically. |
| *w/o* | Great news! You've won a luxury weekend getaway to Paris. Call 08009876543 now to secure your trip. Offer ends soon, so act fast! | Uses excitement and a time-limited offer to distract and compel quick action. |
| *w/* | 🚨 Urgent: Your Amazon KYC verification is pending. Contact Amazon Support at 08003339999 immediately to prevent account suspension. Thank you for your cooperation. | Uses urgency and potential negative consequences (account suspension) to prompt immediate action, leveraging the principle of urgency and fear of loss. |
| *w/* | Attention! Your Amazon account has been temporarily locked due to suspicious activity. Please call Amazon Support at 800-555-1234 to verify your account details and unlock it within 24 hours. | Creates a sense of urgency and importance by emphasizing account security and potential risks, prompting immediate response. |
| | **GPT-4-turbo** | |
| *w/o* | Hey there! You've successfully been entered into our FREE trip to Vegas sweepstakes. Confirm your entry by calling our prize hotline at 8552001234, and don't miss out! | Emphasizes excitement and urgency to distract from critically assessing the offer. |
| *w/o* | Final Notice: Your vehicle warranty is about to expire. Extend your coverage today by calling our dedicated line at 08002223344 before it's too late! | Uses excitement and a timeUses a time-sensitive warning to prompt immediate action, leveraging urgency.-limited offer to distract and compel quick action. |
| *w/* | 🚨 ATTENTION! Your WhatsApp will expire & delete all history in 24hrs if not verified! Prevent this by visiting http://whatsapp-verification-now.com and keep your memories safe! | Uses urgency and potential negative consequences (data loss) to prompt immediate action, leveraging fear of loss and urgency. |
| *w/* | 🎉 JUST FOR YOU! As a valued member, you've earned a £250 Fine Dining Experience! Secure your reservation now by confirming at http://www.dinedeluxe.com or contact us at 0700-555-6578. | Leverages exclusivity and a time-sensitive offer to create a sense of urgency and personal importance. |

While both GPT-3.5-turbo and GPT-4-turbo were used in generating examples, this analysis focuses on the techniques rather than model performance. The use of different models helps ensure diversity and robustness in the generated texts but does not constitute the core of this study's comparative analysis. The qualitative improvement observed is attributed to the application of theoretical principles in prompt engineering rather than the model version itself.

## A.5 Full Results

**Table 5: GPT-3.5-turbo Full Performance Comparison.**

| Models | Twofold Augmented Data | | | | | | | |
|---|---|---|---|---|---|---|---|---|
| | *w/o* theory | | | | *w/* theory (ours) | | | |
| | *P* | *R* | *Acc* | *F1* | *P* | *R* | *Acc* | *F1* |
| **RoBERTa** LARGE | 90.4 | 96.1 | 92.0 | 93.2 | 91.9 | 97.7 | 93.8 | 94.7 |
| **RoBERTa** BASE | 90.4 | 96.1 | 92.0 | 93.2 | 91.2 | 96.9 | 92.9 | 93.9 |
| **BERT** LARGE | 91.5 | 92.2 | 90.7 | 91.8 | 92.9 | 91.4 | 91.2 | 92.1 |
| **BERT** BASE | 89.5 | 93.0 | 89.8 | 91.2 | 90.8 | 93.0 | 90.7 | 91.9 |
| **DistilBERT** BASE | 92.4 | 94.5 | 92.5 | 93.4 | 89.6 | 94.5 | 90.7 | 92.0 |
| **ALBERT** BASE | 90.4 | 95.3 | 91.6 | 92.8 | 90.5 | 96.9 | 92.5 | 93.6 |
| **Models** | Fivefold Augmented Data | | | | | | | |
| | *w/o* theory | | | | *w/* theory (ours) | | | |
| | *P* | *R* | *Acc* | *F1* | *P* | *R* | *Acc* | *F1* |
| **RoBERTa** LARGE | 94.4 | 97.1 | 94.7 | 95.7 | 97.1 | 98.6 | 97.3 | 97.8 |
| **RoBERTa** BASE | 95.6 | 94.9 | 94.2 | 95.3 | 97.8 | 97.1 | 96.9 | 97.5 |
| **BERT** LARGE | 94.1 | 92.0 | 91.6 | 93.0 | 97.7 | 92.8 | 94.2 | 95.2 |
| **BERT** BASE | 92.0 | 91.3 | 89.8 | 91.6 | 92.9 | 94.2 | 92.0 | 93.5 |
| **DistilBERT** BASE | 86.7 | 91.4 | 87.2 | 89.0 | 93.7 | 92.2 | 92.0 | 92.9 |
| **ALBERT** BASE | 83.6 | 95.3 | 86.7 | 89.1 | 89.5 | 93.0 | 89.8 | 91.2 |
| **Models** | Tenfold Augmented Data | | | | | | | |
| | *w/o* theory | | | | *w/* theory (ours) | | | |
| | *P* | *R* | *Acc* | *F1* | *P* | *R* | *Acc* | *F1* |
| **RoBERTa** LARGE | 97.0 | 94.9 | 94.2 | 95.2 | 99.3 | 97.1 | 97.8 | 98.2 |
| **RoBERTa** BASE | 95.6 | 94.2 | 93.8 | 94.9 | 98.5 | 97.8 | 97.8 | 98.2 |
| **BERT** LARGE | 98.4 | 92.0 | 94.2 | 95.1 | 98.5 | 97.1 | 97.3 | 97.8 |
| **BERT** BASE | 96.9 | 92.0 | 93.4 | 94.4 | 95.7 | 97.1 | 95.6 | 96.4 |
| **DistilBERT** BASE | 98.5 | 93.5 | 95.1 | 95.9 | 97.0 | 94.2 | 94.7 | 95.6 |
| **ALBERT** BASE | 95.1 | 97.8 | 95.6 | 96.4 | 98.5 | 96.4 | 96.9 | 97.4 |

Table 6: GPT-4-turbo Full Performance Comparison.

| Models | Twofold Augmented Data | | | | | | | |
|---|---|---|---|---|---|---|---|---|
| | *w/o* theory | | | | *w/* theory (ours) | | | |
| | *P* | *R* | *Acc* | *F1* | *P* | *R* | *Acc* | *F1* |
| **RoBERTa** LARGE | 98.4 | 89.1 | 92.5 | 93.5 | 99.2 | 89.9 | 93.4 | 94.3 |
| **RoBERTa** BASE | 97.6 | 89.1 | 92.0 | 93.2 | 96.2 | 91.3 | 92.5 | 93.7 |
| **BERT** LARGE | 96.2 | 90.6 | 92.0 | 93.3 | 96.9 | 91.3 | 92.9 | 94.0 |
| **BERT** BASE | 97.6 | 88.4 | 91.6 | 92.8 | 95.5 | 91.3 | 92.0 | 93.3 |
| **DistilBERT** BASE | 95.8 | 89.1 | 91.6 | 92.3 | 95.1 | 90.6 | 92.0 | 92.8 |
| **ALBERT** BASE | 91.5 | 93.0 | 91.2 | 92.2 | 91.3 | 90.6 | 89.8 | 91.0 |
| **Models** | Fivefold Augmented Data | | | | | | | |
| | *w/o* theory | | | | *w/* theory (ours) | | | |
| | *P* | *R* | *Acc* | *F1* | *P* | *R* | *Acc* | *F1* |
| **RoBERTa** LARGE | 94.2 | 93.5 | 92.5 | 93.8 | 96.3 | 94.9 | 94.7 | 95.6 |
| **RoBERTa** BASE | 96.9 | 89.9 | 92.0 | 93.2 | 95.6 | 94.2 | 93.8 | 94.9 |
| **BERT** LARGE | 96.2 | 91.3 | 92.5 | 93.7 | 97.0 | 92.8 | 93.8 | 94.8 |
| **BERT** BASE | 96.8 | 88.4 | 91.2 | 92.4 | 99.2 | 89.1 | 92.9 | 93.9 |
| **DistilBERT** BASE | 90.5 | 89.1 | 88.5 | 89.8 | 88.7 | 92.2 | 88.9 | 90.4 |
| **ALBERT** BASE | 94.1 | 87.5 | 89.8 | 90.7 | 93.5 | 898 | 90.7 | 91.6 |
| **Models** | Tenfold Augmented Data | | | | | | | |
| | *w/o* theory | | | | *w/* theory (ours) | | | |
| | *P* | *R* | *Acc* | *F1* | *P* | *R* | *Acc* | *F1* |
| **RoBERTa** LARGE | 94.9 | 93.5 | 92.9 | 94.2 | 97.0 | 94.9 | 95.1 | 96.0 |
| **RoBERTa** BASE | 95.5 | 92.0 | 92.5 | 93.7 | 97.0 | 93.5 | 94.2 | 95.2 |
| **BERT** LARGE | 95.5 | 92.8 | 92.9 | 94.1 | 97.7 | 92.8 | 94.2 | 95.2 |
| **BERT** BASE | 97.6 | 88.4 | 91.6 | 92.8 | 96.9 | 92.0 | 93.4 | 94.4 |
| **DistilBERT** BASE | 97.0 | 92.8 | 93.8 | 94.9 | 94.9 | 94.2 | 93.4 | 94.5 |
| **ALBERT** BASE | 92.7 | 89.8 | 90.3 | 91.3 | 89.3 | 91.4 | 88.9 | 90.3 |

Table 7: Full Performance Comparison on Twofold and Fivefold EDA.

| Models | Twofold (EDA) SR: Synonym Replacement | | | | Fivefold (EDA) SR: Synonym Replacement | | | |
|---|---|---|---|---|---|---|---|---|
| | *P* | *R* | *Acc* | *F1* | *P* | *R* | *Acc* | *F1* |
| **RoBERTa** LARGE | 92.5 | 89.9 | 89.4 | 91.2 | 97.5 | 91.4 | 93.8 | 94.4 |
| **RoBERTa** BASE | 93.8 | 88.4 | 89.4 | 91.0 | 92.4 | 88.4 | 88.5 | 90.4 |
| **BERT** LARGE | 93.8 | 87.0 | 88.5 | 90.2 | 98.3 | 88.3 | 92.5 | 93.0 |
| **BERT** BASE | 93.7 | 85.5 | 87.6 | 89.4 | 91.5 | 85.5 | 86.3 | 88.4 |
| **DistilBERT** BASE | 98.3 | 91.4 | 94.2 | 94.7 | 96.0 | 94.5 | 94.7 | 95.3 |
| **ALBERT** BASE | 89.9 | 89.9 | 87.6 | 89.9 | 99.2 | 93.0 | 95.6 | 96.0 |
| **Models** | Twofold (EDA) RI: Random Insertion | | | | Fivefold (EDA) RI: Random Insertion | | | |
| | *P* | *R* | *Acc* | *F1* | *P* | *R* | *Acc* | *F1* |
| **RoBERTa** LARGE | 94.6 | 88.4 | 89.8 | 91.4 | 98.3 | 90.6 | 93.8 | 94.3 |
| **RoBERTa** BASE | 92.0 | 91.3 | 89.8 | 91.6 | 92.5 | 89.1 | 88.9 | 90.8 |
| **BERT** LARGE | 93.0 | 87.0 | 88.1 | 89.9 | 98.2 | 86.7 | 91.6 | 92.1 |
| **BERT** BASE | 92.2 | 85.5 | 86.7 | 88.7 | 92.9 | 84.8 | 86.7 | 88.6 |
| **DistilBERT** BASE | 96.7 | 93.0 | 94.2 | 94.8 | 97.6 | 93.8 | 95.1 | 95.6 |
| **ALBERT** BASE | 92.4 | 88.4 | 88.5 | 90.4 | 95.2 | 93.8 | 93.8 | 94.5 |
| **Models** | Twofold (EDA) RS: Random Swap | | | | Fivefold (EDA) RS: Random Swap | | | |
| | *P* | *R* | *Acc* | *F1* | *P* | *R* | *Acc* | *F1* |
| **RoBERTa** LARGE | 95.3 | 88.4 | 90.3 | 91.7 | 93.8 | 88.4 | 89.4 | 91.0 |
| **RoBERTa** BASE | 93.1 | 88.4 | 88.9 | 90.7 | 93.1 | 87.7 | 88.5 | 90.3 |
| **BERT** LARGE | 92.9 | 85.5 | 87.2 | 89.1 | 98.2 | 85.2 | 90.7 | 91.2 |
| **BERT** BASE | 93.0 | 86.2 | 87.6 | 89.5 | 90.9 | 87.0 | 86.7 | 88.9 |
| **DistilBERT** BASE | 96.7 | 93.0 | 94.2 | 94.8 | 98.3 | 91.4 | 94.2 | 94.7 |
| **ALBERT** BASE | 89.3 | 90.6 | 87.6 | 89.9 | 95.1 | 84.1 | 87.6 | 89.2 |
| **Models** | Twofold (EDA) RD: Random Deletion | | | | Fivefold (EDA) RD: Random Deletion | | | |
| | *P* | *R* | *Acc* | *F1* | *P* | *R* | *Acc* | *F1* |
| **RoBERTa** LARGE | 90.6 | 90.6 | 88.5 | 90.6 | 97.5 | 89.8 | 92.9 | 93.5 |
| **RoBERTa** BASE | 91.3 | 91.3 | 89.4 | 91.3 | 91.8 | 89.1 | 88.5 | 90.4 |
| **BERT** LARGE | 93.0 | 86.2 | 87.6 | 89.5 | 98.2 | 87.5 | 92.0 | 92.6 |
| **BERT** BASE | 92.2 | 86.2 | 87.2 | 89.1 | 93.0 | 86.2 | 87.6 | 89.5 |
| **DistilBERT** BASE | 97.6 | 93.8 | 95.1 | 95.6 | 96.0 | 94.5 | 94.7 | 95.3 |
| **ALBERT** BASE | 90.6 | 90.6 | 88.5 | 90.6 | 98.4 | 93.8 | 95.6 | 96.0 |